\documentclass[final,1p,times]{elsarticle}
\date{\today}

\usepackage{graphics}
\usepackage{amsmath}

\biboptions{square,sort&compress}

\begin{document}

\title{Shell evolution in neutron-rich carbon isotopes: Unexpected enhanced role of neutron-neutron correlation}

\author[rvt]{C.X. Yuan}

\author[rvt,focal]{C. Qi}

\author[rvt,els]{F.R. Xu\corref{cor1}}
\ead{frxu@pku.edu.cn}

\cortext[cor1]{Corresponding author at: School of Physics, Peking
University, Beijing 100871, China}

\address[rvt]{State Key Laboratory of Nuclear Physics and Technology,
School of Physics, Peking University, Beijing 100871, China}
\address[focal]{KTH (Royal Institute of Technology), Alba Nova
University Center, SE-10691 Stockholm, Sweden}
\address[els]{Center for Theoretical Nuclear Physics, National
Laboratory for Heavy Ion Physics, Lanzhou 730000, China}

\begin{abstract}
Full shell-model diagonalization has been performed to study the
structure of neutron-rich nuclei around $^{20}$C. We investigate in
detail the roles played by the different monopole components of the
effective interaction in the evolution of the $N=14$ shell in C, N
and O isotopes. It is found that the relevant neutron-neutron
monopole terms, $V^{nn}_{d_{5/2}d_{5/2}}$ and
$V^{nn}_{s_{1/2}s_{1/2}}$, contribute significantly to the reduction
of the $N=14$ shell gap in C and N isotopes in comparison with that
in O isotopes. The origin of this unexpectedly large effect, which
is comparable with (sometimes even larger than) that caused by the
proton-neutron interaction, is related to the enhanced configuration
mixing in those nuclei due to many-body correlations. Such a scheme
is also supported by the large $B(E2)$ value in the nucleus $^{20}$C
which has been measured recently.
\end{abstract}


\maketitle

\section{\label{sec:level1}Introduction}
The study of neutron-rich nuclei with unusually large $N/Z$ ratios
is challenging the conventional view of nuclear
structure~\cite{solin2008,janssens2009}. It is established that,
when going from the $\beta$-stability line to drip line, the shell
structure evolves and new magic numbers may emerge due to the
dynamic effects of the nucleon-nucleon interaction. The influence of
the proton-neutron interaction and its higher order term (namely the
tensor force) has been extensively analyzed in recent
publications~\cite{otsuka2001,otsuka2005,otsuka2010,solin2008,Smi10,Uts12,umeya2006}.
One may expect that other components of the nucleon-nucleon
interaction contribute also to the shell evolution. However, this
issue has not been much covered, which may be related to the fact
that the isoscalar channel of the effective interaction is usually
much stronger than the isovector part. In Ref.~\cite{suzuki2009}, it
was found that a modification for the isovector $T=1$ channel of the
monopole interaction is needed to reproduce the energies and
electromagnetic transition properties of low-lying states in nuclei
around $^{17}$C.

In this paper we analyze in detail contributions from the different
terms of monopole interaction to the evolution of shell structure.
In particular, we will show that the neutron-neutron monopole
interaction can have significant effect on the evolution of $N=14$
shell in carbon isotopes. The rapid decrease of the $E(2^{+}_{1})$
in $N=14$ isotones indicates that the $N=14$ shell gap existing in
oxygen isotopes erodes in nitrogen and carbon
isotopes~\cite{stanoiu2004,stanoiu2008,sohler2008,solin2008,strongman2009,elekes2010}.
At $N=14$, the $B(E2)$ value measured through the lifetime of the
$2^{+}_{1}$ state in $^{20}$C~\cite{petri2011} is much larger than
the values in $^{16,18}$C~\cite{wiedeking2008,ong2008} and
$^{22}$O~\cite{raman2001} while an inelastic scattering measurement
presents a much smaller $B(E2)$ value in $^{20}$C~\cite{elekes2009}.
In this paper, we will also analyze $E2$ transition properties in
neutron-rich carbon isotopes from the viewpoint of collectivity.

In Sec.~\ref{sec:level2} we briefly introduce the theoretical
framework. The influence of different monopole terms of the
effective interaction on the evolution of the $N=14$ shell is
discussed in Sec.~\ref{sec:level3}. In Sec.~\ref{sec:level4} we
analyze $E2$ decay properties of the nuclei $^{16, 18, 20}$C. The
calculations are summarized in Sec.~\ref{sec:level5}.

\section{\label{sec:level2}Theoretical framework}
The model space we choose contains $p$ and $sd$ shells~\cite{brown}.
To facilitate the calculation, we restrict the maximum number of two
for nucleons that can be excited from $p$ to $sd$ shell. This
truncation is denoted as $2\hbar\omega$ space as usual. It should be
mentioned that the full  $2\hbar\omega$ space should also include
contributions such as the excitation of one nucleon from the $p$
shell to the $fp$ shell or from the $0s$ shell to the $sd$ shell.
Such configurations are neglected in the present study, since they
do not show any significant influence on the structure of nuclei of
concern. Similarly, in the present work, the $0\hbar\omega$
truncation implies that no nucleon is excited from the $p$ shell to
the $sd$ shell. The well-established WBT~\cite{WBT92} and
MK~\cite{MK} effective Hamiltonians are used. The WBT interaction is
constructed based on the USD~\cite{usd} interaction. The $p$-shell
and $psd$-cross-shell matrices in WBT are determined by fitting
experimental data. The $sd$, $p$, $psd$ and $ppsdsd$ parts in the MK
interaction are taken from the PW interaction~\cite{pw}, CK
interaction~\cite{ck}, MK potential~\cite{MK} and Kuo-Brown G
matrix~\cite{gmatrix}, respectively. Calculations are carried out
with a newly-established parallel shell-model code described in
Ref.~\cite{Qi07}.

In $2\hbar\omega$ calculations, one has to remove the spurious
states which are caused by the center-of-mass motion. As suggested
in Ref.~\cite{cm1974}, the shell model Hamiltonian $H$ is modified
to be $H'=H+\beta H_{c.m.}$ where $H_{c.m.}$ is the center-of-mass
Hamiltonian. One can separate the low-lying states from the spurious
states by taking a large positive $\beta$ value, which can move the
center-of-mass excitations to high excitation energies. We take
$\beta=100$~MeV in the present work, which is used in some previous
works in this region~\cite{umeya20082,ma2010,utsuno2011}.

The effective single-particle energy (ESPE)~\cite{otsuka20012},
describes in a simple manner the evolution of shell structure within
the shell model
framework~\cite{otsuka2001,otsuka2005,otsuka2010,solin2008,Smi10,umeya2006}.
The ESPE of a given $j$ orbit is written
as~\cite{umeya2006,otsuka20012}
\begin{equation}\label{speequ}
 \varepsilon_{j}=\varepsilon_{j}^{core}+\sum_{j'}V_{jj'}\langle\psi|\widehat{N}_{j'}|\psi\rangle,
\end{equation}
where $\varepsilon_{j}^{\rm core}$ is the single-particle energy
with respect to the core, $\langle\psi|\widehat{N}_{j'}|\psi\rangle$
is the occupancy of nucleons in the $j'$ orbit and $V_{jj'}$ is the
monopole interaction \cite{Ban}. As in Ref.~\cite{umeya2006}, one
has
\begin{equation}\label{monople}
V_{jj',T} =
\sum_{J}\left[1-(-1)^{(2j-J-T+1)}\delta(jj')\right]\frac{(2J+1)}{(2j+1)(2j'+1)}\langle
jj'|V|jj'\rangle_{JT},
\end{equation}
where $\langle jj'|V|jj'\rangle_{JT}$ is the two-body matrix element
and $J$ and $T$ are the angular momentum and the isospin of the
corresponding two-particle state, respectively. To explore the
influence of configuration mixing on the ESPE, we will adopt the
occupancy in Eq.~(\ref{speequ}) obtained from the full shell-model
diagonalization.

\section{Calculations and discussions}

\subsection{\label{sec:level3}N=14 shell closure}

In the independent particle model, the last neutrons in $^{16}$C,
$^{18}$C and $^{20}$C should occupy the $\nu0d_{5/2}$ orbit as in O
isotopes. However, the ground state and the first excited state of
$^{15}$C are $1/2^{+}$ and $5/2^{+}$ states, respectively, which
indicates that the $\nu1s_{1/2}$ and $\nu0d_{5/2}$ orbits are
inverse in $^{15}$C with respect to $^{17}$O. As noted in
Ref.~\cite{zheng2002}, $^{16}$C shows a strong $(\nu1s_{1/2})^{2}$
configuration. The rapid decrease of $E(2^{+}_{1})$ in $N=14$
isotones from O to C~\cite{stanoiu2004,sohler2008,stanoiu2008} also
indicates that shell structures in O and C isotopes are different.

\begin{figure*}
\includegraphics[scale=0.5]{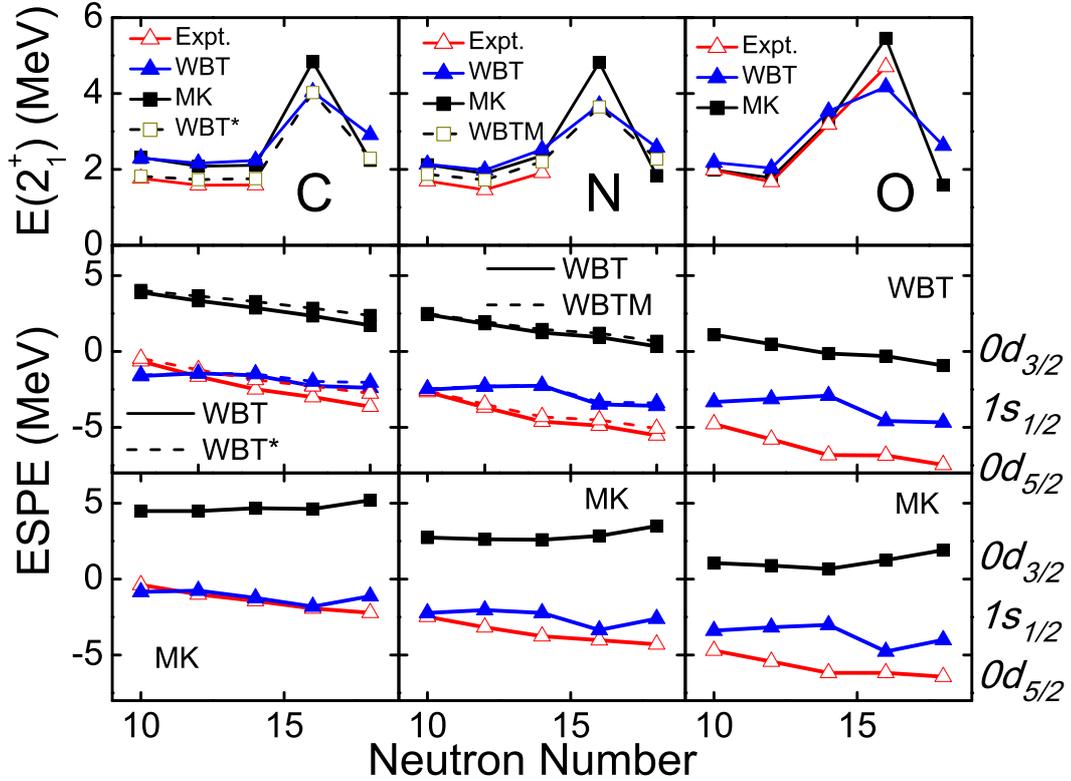}
\caption{\label{spegra} Calculated $E(2^{+}_{1})$ and ESPEs in
even-neutron C, N and O isotopes. In N isotopes, the effective
$E(2^{+}_{1})$ are deduced from the $2J+1$ weighted average of
$E(3/2^{+}_{1})$ and $E(5/2^{+}_{1})$~\cite{sohler2008}.
Experimental data are from
Refs.~\cite{raman2001,stanoiu2004,stanoiu2008,sohler2008,hoffman2009}.}
\end{figure*}

In order to study the structure difference between O and C isotopes,
firstly we present $E(2^{+}_{1})$ and ESPEs in the even-neutron C, N
and O isotopes, as shown in Fig.~\ref{spegra}. For N isotopes, the
energies are effective $E(2^{+}_{1})$ which are deduced from the
$2J+1$ weighted average of $E(3/2^{+}_{1})$ and
$E(5/2^{+}_{1})$~\cite{sohler2008}. It is seen that the calculated
$E(2^{+}_{1})$ values with WBT and MK interactions are
systematically larger than experimental values in C and N isotopes.
It has been pointed out that a reduction of $V^{nn}$ in the $sd$
part of WBT can improve the results~\cite{stanoiu2008,sohler2008}.
For such an improvement, the modified versions of WBT have been
suggested by multiplying the $sd$ part of $V^{nn}$ in WBT by a
factor of $0.75$ for C isotopes (named WBT*~\cite{stanoiu2008}) and
$0.875$ for N isotopes (named WBTM~\cite{sohler2008}). We can see
that the $E(2^{+}_{1})$ calculations are improved with WBT* for C
isotopes and WBTM for N isotopes, shown in Fig.~\ref{spegra}. From
$^{18}$O to $^{16}$C, the quenching of $V^{nn}$ is explained via the
core polarization effect~\cite{sieja2011}. In Fig.~\ref{spegra}, we
do not show the $2\hbar\omega$ results, since they are very similar
to the $0\hbar\omega$ calculations. Both the energy and the
$\nu0d_{5/2}-\nu1s_{1/2}$ gap in N isotopes are almost in the middle
of those in C and O isotopes.

In C isotopes, the calculated energy gap between $\nu1s_{1/2}$ and
$\nu0d_{5/2}$ orbits is much smaller than what is shown in
Ref.~\cite{stanoiu2008}. By calculating ESPEs with shell-model
occupancy, the correlation effect is approximately included. So the
ESPEs are different from those calculated with simple occupancy like
assuming six neutrons all occupying $\nu0d_{5/2}$ orbits in $^{20}$C
and $^{22}$O. The WBT* and WBTM give smaller
$\nu0d_{5/2}-\nu1s_{1/2}$ gaps for C and N isotopes than those by
WBT.

The role that each interaction plays in evolution of $N=14$ shell
can be investigated based on present ESPEs. The energy gap between
$\nu1s_{1/2}$ and $\nu0d_{5/2}$ is written as
\begin{eqnarray}
  \Delta\varepsilon_{N=14} &=&
  \varepsilon_{\nu1s_{1/2}}-\varepsilon_{\nu0d_{5/2}}.
\end{eqnarray}
The difference between $^{20}$C and $^{22}$O is written as
\begin{eqnarray}
  \Delta\varepsilon_{N=14}(\mbox{$^{22}$O$-^{20}$C}) &=&
 \Delta\varepsilon_{N=14}(^{22}\mbox{O})-\Delta\varepsilon_{N=14}(^{20}\mbox{C})
  \nonumber\\
  &=&
  [\varepsilon_{\nu1s_{1/2}}(^{22}\mbox{O})-\varepsilon_{\nu1s_{1/2}}(^{20}\mbox{C})]
  -[\varepsilon_{\nu0d_{5/2}}(^{22}\mbox{O})-\varepsilon_{\nu0d_{5/2}}(^{20}\mbox{C})]
  \nonumber\\
  &=&
  \sum_{j'}[\langle\nu1s_{1/2}j'|V|\nu1s_{1/2}j'\rangle_{JT}-\langle\nu0d_{5/2}j'|V|\nu0d_{5/2}j'\rangle_{JT}]\times
  \nonumber\\
  & &[N_{j'}(^{22}\text{O})-N_{j'}(^{20}\mbox{C})]\label{deltae}.
\end{eqnarray}

The first part is the monopole interaction between the $j'$-th orbit
and $\nu1s_{1/2}$ ($\nu0d_{5/2}$) orbit, while the second part is
the difference of occupancy on $j'$-th orbit between $^{22}$O and
$^{20}$C. Besides two $0p_{1/2}$ protons, the numbers of protons and
neutrons on other orbits are also different between $^{22}$O and
$^{20}$C. The contribution from each term of the monopole
interaction can be analyzed by Eq.~(\ref{deltae}). There are totally
$20$ terms of the monopole interaction which contribute in
$2\hbar\omega$ calculation. In $0\hbar\omega$ calculation, the
number of the terms reduces to $10$ because no protons are in $sd$
shell and neutrons in $p$ shell are fully occupied in $^{22}$O and
$^{20}$C.

\begin{figure}
\includegraphics[scale=0.5]{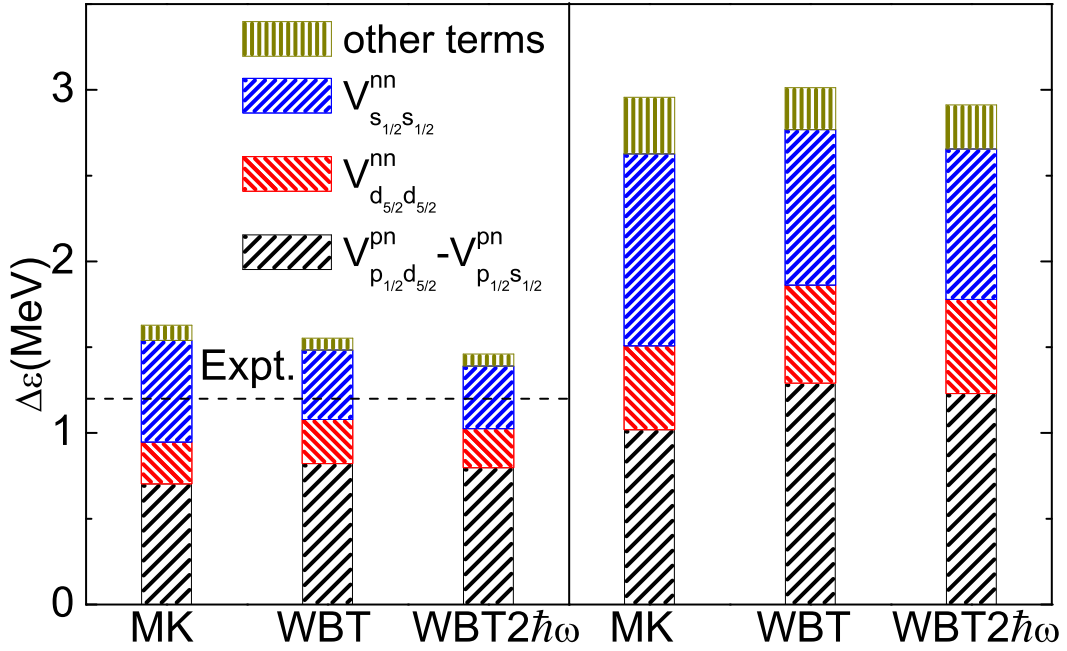}
\caption{\label{gapONC} Contribution of each interaction in $N=14$
gap difference between $^{22}$O and $^{21}$N (left panel), $^{22}$O
and $^{20}$C (right panel). The dashed line is for the value
obtained from observed levels in $^{22}$O and
$^{21}$N~\cite{sohler2008}. The $2\hbar\omega$ calculations are
indicated, otherwise $0\hbar\omega$ calculations.}
\end{figure}

Fig.~\ref{gapONC} presents
$\Delta\varepsilon_{N=14}(\mbox{$^{22}$O$-^{21}$N})$ and
$\Delta\varepsilon_{N=14}(\mbox{$^{22}$O$-^{20}$C})$, respectively.
It is seen that the proton-neutron interaction,
$V^{pn}_{p_{1/2}d_{5/2}}-V^{pn}_{p_{1/2}s_{1/2}}$, has large
contribution to $\Delta\varepsilon_{N=14}$ as discussed in
Ref.~\cite{stanoiu2008}. The two terms, $V^{nn}_{d_{5/2}d_{5/2}}$
and $V^{nn}_{s_{1/2}s_{1/2}}$, have also significant contributions
to $\Delta\varepsilon_{N=14}$. In $2\hbar\omega$ calculation, the
influence of other $16$ terms is small. When removing two protons
from $^{22}$O, the proton-neutron interaction reduces the $N=14$ gap
and enlarges the mixing between $\nu1s_{1/2}$ and $\nu0d_{5/2}$
orbits. At the same time, the neutron-neutron interaction
contributes to the shell evolution because the neutrons in
$\nu1s_{1/2}$ and $\nu0d_{5/2}$ orbits rearrange as a result of
correlation effect from shell-model diagonalization. Although the
ESPEs do not entirely include the correlation effect because the
multipole interaction is not taken into account, we can
approximately study the role of neutron-neutron correlation in shell
evolution by applying occupancy obtained from full shell-model
diagonalization to Eq.~(\ref{speequ}). The attractive
neutron-neutron interactions are mostly due to the attractive
singlet-even (SE) channel of the central force~\cite{umeya2006}.
Specially, $V^{nn}_{s_{1/2}s_{1/2}}$ is completely from this
channel.

The study of shell evolution extends our understanding of
nucleon-nucleon interaction, especially the tensor
force~\cite{otsuka2001,otsuka2005,otsuka2010,solin2008,Smi10,Uts12,umeya2006}.
From a general view of shell evolution, when proton (neutron) number
changes in isotones (isotopes), the neutron (proton) shell structure
evolves due to the proton-neutron interaction. Here we present a
detailed investigation on $N=14$ shell evolution to point out that
the role played by the proton-neutron interaction in shell evolution
can be ``enhanced'' by the attractive neutron-neutron interaction
due to the many-body correlations. It is important to notice this
``enhancement'' if one wants to study the nuclear force from shell
evolution. It is interesting to investigate the contribution of
$V^{pp}_{jj}$ and $V^{nn}_{jj}$ in other cases of shell evolution in
order to gain a complete understanding of how nuclear force drives
the shell evolution. The $N=16$ shell existed in O isotopes will
dramatically shrink in Si isotopes~\cite{otsuka2001}. We adopt the
similar method which we use in the study of $N=14$ shell for a
detailed investigation of the $N=16$ shell evolution from $^{24}$O
to $^{30}$Si. The result shows that, besides the proton-neutron
interaction which was discussed in Ref.~\cite{otsuka2001}, the
neutron-neutron interactions, $V^{nn}_{d_{3/2}d_{3/2}}$ and
$V^{nn}_{s_{1/2}s_{1/2}}$, contribute to the $N=16$ shell evolution
from $^{24}$O to $^{30}$Si. Because of the limitation of
calculation, it is hard to perform full shell-model diagonalization
to study shell evolution in heavier regions, such as that in $N=51$
isotones which was discussed in Refs.~\cite{otsuka2005,otsuka2010}.
One may expect that more information from shell evolution can be
gained through full shell-model diagonalization in heavier regions
in the future.

\subsection{\label{sec:level4}$E2$ decay properties}

Existing shell-model interactions reproduce well the observed
$B(E2)$ values for carbon isotopes~\cite{suzuki2008,petri2011}
including the large $B(E2)$ value in $^{20}$C. In shell model, the
$B(E2)$ is calculated by
\begin{equation}
B(E2,j_{i}\rightarrow
j_{f})=\frac{1}{2j_{i}+1}[e_{p}^{eff}A_{p}+e_{n}^{eff}A_{n}]^{2},
\end{equation}
where $j_{i}$ and $j_{f}$ are the angular moments of the initial and
final states of the $E2$ transition. $A_{p}$ and $A_{n}$ are proton
and neutron $E2$ transition matrix elements, respectively. In
Table~\ref{BE2}, we present $A_{p}$, $A_{n}$ and $B(E2)$ values for
even-even $^{16-20}$C. In the calculations, the oscillator parameter
of $\hbar\omega=45A^{-1/3}-25A^{-2/3}$ is employed. An approximate
$1/A$-dependent effective charge is used~\cite{sagawa2004}. For
$^{16-20}$C, the effective charges are $e_{p}^{eff}(e_{n}^{eff})$
$=$ $1.16(0.33)$, $1.11(0.27)$, $1.07(0.22)$,
respectively~\cite{sagawa2004}.
 It is seen that the $A_{p}$ increases significantly from $^{16}$C
to $^{20}$C. Its contribution in $B(E2)$ values is multiplied by
$e_{p}^{eff}$. Thus the  $B(E2)$ value in $^{20}$C is much larger
than those in $^{16}$C and $^{18}$C. As discussed in
Refs.~\cite{suzuki2008,petri2011}, the reduction in the
$\pi0p_{1/2}-\pi0p_{3/2}$ gap from $^{16}$C to $^{20}$C would be one
reason why the proton excitation probability and the $B(E2)$ value
increase from $^{16}$C to $^{20}$C.

\begin{table}
\begin{center}
\caption{\label{BE2} Calculated $E2$ transition matrix elements and
$B(E2)$ values for even-even $^{16-20}$C. The SFO and SFO-tls
calculations~\cite{suzuki2008} are given for comparison.}
\vspace{1mm} \setlength{\tabcolsep}{20pt}
\begin{tabular}{cccc}
\hline
           &    $A_{p}~(fm)$  &  $A_{n}~(fm)$  &   $B(E2)~(e^{4}fm^{2})$     \\
\hline
$^{16}$C                  &      &         & \\
MK-$0\hbar\omega$          &    1.97 & 8.96 &  27.52    \\
WBT-$0\hbar\omega$         &    1.27 & 9.39 &  20.88    \\
WBT-$2\hbar\omega$         &    1.12 & 9.40 &  19.40    \\
WBT*-$0\hbar\omega$        &    1.16 & 9.31 &  19.55    \\
SFO                       &    1.20 & 8.27 &  17.00    \\
SFO-tls                   &    1.38 & 8.06 &  18.20    \\
Expt.                     &         &     &        13.0$\pm$1.0$\pm$3.5~\cite{ong2008}    \\
                          &         &     &      20.8$\pm$3.7~\cite{wiedeking2008}      \\
$^{18}$C                  &      &         &     \\
MK-$0\hbar\omega$          &2.25 &  11.22 & 30.55        \\
WBT-$0\hbar\omega$         &1.77 &  11.16 & 24.74        \\
WBT-$2\hbar\omega$         &1.78 &  11.15 & 24.89        \\
WBT*-$0\hbar\omega$        &1.78 &  11.08 & 24.69        \\
SFO                       &1.58 &  9.87  & 19.50        \\
SFO-tls                   &1.60 &  9.69  & 19.90        \\
Expt.                     &     &        &       21.5$\pm$1.0$\pm$5.0~\cite{ong2008}        \\
\\
$^{20}$C                  &      &         &       \\
MK-$0\hbar\omega$          & 3.64 &  11.82  &42.19        \\
WBT-$0\hbar\omega$         & 3.07 &  11.50  &33.85        \\
WBT-$2\hbar\omega$         & 3.00 &  11.32  &32.43        \\
WBT*-$0\hbar\omega$        & 2.96 &  11.30  &31.95        \\
Expt.                     &     &          &37.5$^{+15.0+5.0}_{-8.5-2.0}$~\cite{petri2011} \\
\hline
\end{tabular}
\end{center}
\end{table}

The large $B(E2)$ value in $^{20}$C can also be explained from the
view of collectivity. In our previous work, we claimed that $^{20}$C
may have large quadrupole deformation~\cite{yuan09}. The intrinsic
quadrupole moment $Q_{0}(s)$ of the $2_{1}^{+}$ state can be deduced
from the spectroscopic quadrupole moment $Q_{spec}$. If the
$2_{1}^{+}$ state is a rotational deformed state, the intrinsic
quadrupole moment (written as $Q_{0}(t)$) can also be deduced from
the $B(E2;2_{1}^{+} \rightarrow 0_{1}^{+})$ value of the $E2$
transition. In the case of a good deformed rotor, these two
definitions of the intrinsic quadrupole moments should give the
similar values. One has
\begin{eqnarray}
  Q_{0}(s) &=& \frac{(J+1)(2J+3)}{3K^{2}-J(J+1)}Q_{spec}(J),~(K\neq1),
\end{eqnarray}
and
\begin{eqnarray}
B(E2;J \rightarrow J-2)&=&
\frac{5}{16\pi}e^{2}|<JK20|(J-2)K>|^{2}Q_{0}(t)^{2},\nonumber
\\
& &(K\neq1/2,1).
\end{eqnarray}
In our case, we have $K=0$ and $J=2$. The values of $B(E2)$ and
$Q_{spec}$ can be calculated in shell model. Table~\ref{Q} lists the
intrinsic quadrupole moments for $^{16}$C, $^{18}$C and $^{20}$C,
and the ratios between $Q_{0}(s)$ and $Q_{0}(t)$.

\begin{table}
\begin{center}
\caption{\label{Q}Calculated intrinsic quadrupole moments $Q_{0}(s)$
(from spectroscopic property) and $|Q_{0}(t)|$ (from $E2$
transition), and the ratio $|Q_{0}(s)/Q_{0}(t)|$. The experimental
$|Q_{0}(t)|$ are deduced from the observed $B(E2)$
values~\cite{wiedeking2008,ong2008,petri2011}.} \vspace{1mm}
\setlength{\tabcolsep}{20pt}
\begin{tabular}{cccc}
\hline
  &$Q_{0}(s)(fm^{2})$&$|Q_{0}(t)|(fm^{2})$&$|Q_{0}(s)/Q_{0}(t)|$ \\
\hline
$^{16}$C        &   &&
\\
MK-$0\hbar\omega$  & $10.07$ & $16.63$&$0.61$
\\
WBT-$0\hbar\omega$ & $13.25$ &$14.48$&$0.91$
\\
WBT-$2\hbar\omega$ & $13.93$ & $13.96$&$1.00$
\\
WBT*-$0\hbar\omega$& $13.53$ & $14.02$&$0.97$
\\
Expt. &   & $11.43\pm1.98$~\cite{ong2008}&
\\
      &   & $14.44\pm1.27$~\cite{wiedeking2008}&
\\
 $^{18}$C        &   &&
\\
MK-$0\hbar\omega$  & $11.02$ & $17.52$&$0.63$
\\
WBT-$0\hbar\omega$ & $11.19$ & $15.77$&$0.71$
\\
WBT-$2\hbar\omega$ & $8.72$  & $15.81$&$0.55$
\\
WBT*-$0\hbar\omega$& $9.21$ & $15.75$&$0.58$
\\
Expt. &   & $14.70\pm2.05$~\cite{ong2008}&
\\
\\
 $^{20}$C        &   &&
\\
MK-$0\hbar\omega$ & $-20.14$ & $20.59$&$0.98$
\\
WBT-$0\hbar\omega$& $-18.36$ & $18.44$&$1.00$
\\
WBT-$2\hbar\omega$& $-17.55$ & $18.05$&$0.97$
\\
WBT*-$0\hbar\omega$& $-17.44$ & $17.92$&$0.97$
\\
Expt. &   & $19.41^{+5.18}_{-2.72}$~\cite{petri2011}&
\\
 \hline
\end{tabular}
\end{center}
\end{table}

It is seen that $Q_{0}(s)$ and $Q_{0}(t)$ are almost the same in
$^{20}$C, which, associated with a negative quadrupole moment, would
indicate that $^{20}$C may have an oblate deformed structure, and
thus the large $B(E2)$ value in $^{20}$C might be understood with a
consideration of the deformed collectivity. It may be of interest to
understand why such collectivity is found in a nucleus as light as
$^{20}$C. The six valence neutrons have strong configuration mixing
due to the disappearance of the $N=14$ shell, and, due to the
reduction of the proton $0p_{1/2}-0p_{3/2}$ gap in $^{20}$C, the
probability of the proton excitation to $0p_{1/2}$ orbit increases,
as discussed above. The strong mixing of configurations in both
protons and neutrons would induce the enhanced collective property
in $^{20}$C. The results from AMD~\cite{enyo2005} and deformed
Skyrme-Hartree-Fock~\cite{sagawa2004} have also shown an oblate
deformation for $^{20}$C. Compared with $^{20}$C, the ratio between
$Q_{0}(s)$ and $Q_{0}(t)$ in $^{18}$C is significantly less than the
number of $1$ in both WBT and MK calculations. Although we can not
exclude other collective freedoms in $^{18}$C in the present study,
the smaller $Q_{0}(s)$ in $^{18}$C compared with $^{20}$C would
indicate that the quadrupole deformation in $^{18}$C should be
smaller than that in $^{20}$C. In case of $^{16}$C, the WBT and MK
give rather different results. Calculations with the WBT interaction
show that the nucleus $^{16}$C is well prolate deformed.

In the shell model where a spherical basis is used usually, the
collectivity (or deformation) may be discussed in the language of
configuration admixtures. In mean-field models, the collectivity (or
deformation) can be discussed qualitatively by the filling of
deformed orbits. The deformed Skyrme-Hartree-Fock calculation has
been performed for $^{17}$C~\cite{sagawa2008}, giving coexisting
prolate and oblate minima in each potential energy surface of the
ground and lowly-excited states. It was predicted that the $3/2^{+}$
g.s. has the lowest energy minimum at a prolate deformation while
the first excited $1/2^{+}$ state has the lowest minimum at an
oblate shape~\cite{sagawa2008}. The authors explained also that the
observed hindered $M1$ transition between the $1/2^{+}$ and
$3/2^{+}$ states would be attributed to the shape difference of the
two states~\cite{sagawa2008}. If we look at the Nilsson diagram of
single-particle orbits, the $1/2[220]$ orbit of the $0d_{5/2}$ shell
has a prolate-driving force and $5/2[202]$ has an oblate-driving
force, while $3/2[211]$ has a prolate-oblate competing force. For
the $3/2^{+}$ g.s. in $^{17}$C, the three valence neutrons outside
the $N=8$ closed shell would have two on the $1/2[220]$ orbit and
one on $3/2[211]$, which would lead to a prolate deformation for the
ground state. The first excited $1/2^{+}$ state at the lowest energy
should have two neutrons on $5/2[202]$ and one on $1/2[220]$, which
would make an oblate deformation. Back to the even-even C isotopes
which we are investigating, the similar qualitative analyses of
deformations may be made. The $^{16, 28, 20}$C isotopes should have
coexisting prolate and oblate shapes, similar to $^{17}$C. The g.s.
of $^{16}$C would favor a prolate shape with two valence neutrons
occupying the prolate $1/2[220]$ orbit. $^{20}$C would favor an
oblate deformation with two neutrons on the oblate $5/2[202]$ orbit,
while $^{18}$C might be more shape-competitive between prolate and
oblate deformations. These qualitative analyses are consistent with
the shell-model calculations given in Table~\ref{Q}. However, it
should be noted that the deformations of the carbon isotopes are
rather soft, compared to heavy nuclei.

\section{\label{sec:level5}Summary}

In summary, the nuclei around $^{20}$C have been studied with shell
model in $0\hbar\omega$ and $2\hbar\omega$ model spaces. The $N=14$
shell closure appears in O isotopes clearly, but vanishes in C
isotopes~\cite{stanoiu2008,sohler2008}. Besides the proton-neutron
interaction which was discussed in Ref.~\cite{stanoiu2008}, the
neutron-neutron interactions, $V^{nn}_{d_{5/2}d_{5/2}}$ and
$V^{nn}_{s_{1/2}s_{1/2}}$, play also important role. When the $N=14$
shell gap is reduced, neutrons in the $1s_{1/2}$ and $0d_{5/2}$
orbits rearrange and the interactions $V^{nn}_{d_{5/2}d_{5/2}}$ and
$V^{nn}_{s_{1/2}s_{1/2}}$ can have significant effects on the shell
evolution. In $2\hbar\omega$ calculation, other $16$ terms of
monopole interaction have very small contributions to the shell
evolution. It would be useful to do systematic investigations on the
contributions of $V^{pp}_{jj}$ and $V^{nn}_{jj}$ in order to
understand the nuclear force from shell evolution.

As a special case of nuclei near the neutron drip line, we have
analyzed the $B(E2)$ value for $^{20}$C in which experimental data
have been available. In Refs.~\cite{suzuki2008,petri2011}, it was
pointed out that the large $B(E2)$ in $^{20}$C is caused by the
reduction of the proton $0p_{1/2}-0p_{3/2}$ gap, which leads to a
large proton transition matrix element. The present shell-model
calculation gives the similar result. Usually, a large $B(E2)$ value
is connected with collectivity (or deformation). Therefore, we have
made a further investigation of the problem. Our calculation gives a
large quadrupole moment (or large $B(E2)$) for $^{20}$C, indicating
an increased collectivity (or deformation). In the shell-model
language, the enhanced collectivity is caused by strong
configuration mixings in both valence protons and neutrons.

\section*{Acknowledgement}
This work has been supported by the National Natural Science
Foundation of China under Grant No. 10975006. CQ acknowledges the
supports by the Swedish Research Council (VR) under grant Nos.
623-2009-7340 and 621-2010-4723 and the computational support
provided by the Swedish National Infrastructure for Computing (SNIC)
at PDC and NSC.

\bibliographystyle{elsarticle-num}

\end{document}